# Incremental Classification using Feature Tree

N N Vadnere[#1], R G Mehta[*2], D P Rana[*3], N J Mistry[*], M M Raghuwanshi[**]

[#] *M.Tech, Computer Science Department, Manipal University Jaipur, Jaipur, India*
[1] `nishantvadnere@yahoo.com`
[*] *SVNIT, Surat, India*
[2] `rgm@coed.svnit.ac.in`
[3] `dpr@coed.svnit.ac.in`
[**] *RGCET, Nagpur, India*

*Abstract*— **In recent years, stream data have become an immensely growing area of research for the database, computer science and data mining communities. Stream data is an ordered sequence of instances. In many applications of data stream mining data can be read only once or a small number of times using limited computing and storage capabilities. Some of the issues occurred in classifying stream data that have significant impact in algorithm development are size of database, online streaming, high dimensionality and concept drift. The concept drift occurs when the properties of the historical data and target variable change over time abruptly in such a case that the predictions will become inaccurate as time passes. In this paper the framework of incremental classification is proposed to solve the issues for the classification of stream data. The Trie structure based incremental feature tree, Trie structure based incremental FP (Frequent Pattern) growth tree and tree based incremental classification algorithm are introduced in the proposed framework.**

## I. Introduction

In recent years, stream data have become a rising significant area of research for the database, computer science and data mining communities. Stream data is an ordered sequence of instances that in many applications of data stream mining can be read only once or a small number of time using limited computing and storage capabilities. The classification results change over time when concept drift occurs. This leads to change in data patterns. It is also known as data stream evolution. It has some other issues such as size of data, dimensionality and online streaming. Generally the data mining applications involve millions or billions of records with high dimensionality, which make the task of classification computationally very expensive. Recent applications of Data Mining deals with incremental data so whenever the database is updated the reconstruction of the classification model is needed, which involves many complex, time consuming tasks for the entire dataset. The incremental approach in data mining deals with the model preparation and maintenance incrementally for the updated datasets.

In this paper the framework is proposed for incremental classification for high dimensional stream data. As part of process, first step is to prepare the data which include preprocessing techniques like normalization, missing value replacement, transformation, and discretization. The next step is to make feature tree from preprocessed data. The feature tree is proposed using Trie structure to store the detail of the dataset then FP growth tree is created from the feature tree to reduce the feature for the classification process. The tree based dynamic incremental classification is proposed, which keeps on updating to incorporate the concept drift.

The paper is structured as follows. A brief review of related work is given in Section II which includes preprocessing, data mining algorithm for Trie structure and classification. Section III presents proposed framework and finally Section IV contains summary and future scope.

## II. Related Work

To achieve the incremental classification here this section is describing the literature review required at various steps of the proposed framework like data preprocessing, Trie structure and classification.

### A. Data Preprocessing

Data preprocessing is the substantial step in data mining. Data mining mainly involve missing value replacement, transformation, normalization, and discretization. The result of data preprocessing is the final training set.

There are many originators of missing value such as broken sensor, erroneous or missing data entries and in some case some attributes make no sense for some type of objects. It is necessary to replace the missing value otherwise the analysis could lead to meaningless denouement. The clear way is to replace the missing value by mean value in case of numeric attributes and modus in case of categorical value [1]. S. McClean et al. proposed a technique to replace the missing value by making rules based on background knowledge [2]. J. J. Shen et al. recommended Rule Recycle bin technique which restate and construct the rules in order to get more complete attributes value association rule. This will enable the database recovered to advance the accuracy and completion rate as well as gain the validity of missing value completion [3]. M. Shyu et al. designed a framework named F-DCS for replacing missing value which accepts the basic concept of conditional probability theories. This framework can manage both nominal and numeric values with a high degree of accuracy when it is compared with other techniques such as using minimum, average and maximum value [4]. R. Malaryizhi et al. enlisted K-NN classifier that performs better than K-means clustering in missing value imputation [5]. N. Devi et al. projected to replace the missing value by mean and median of clusters and got the more accurate result for their classification analysis for different number of clusters [6].

Normalization is also important task in data preprocessing to reduce unwanted variation either within or between arrays. Normally normalization can be done on data with three ways such as Z-Score, by decimal scaling or min-max normalization. There are main two types of normalization based on a) distance and b) proportion. Distance based normalization includes vector based that is on Euclidian distance and linear based normalization which can correct skewedness in data. Proportion normalization includes non-monotonic normalization which is on Z-Score. The normalization property requires that the range of a sameness or distance measure lies within a fixed range.

Transformation is also a valuable step in data preprocessing. Transformation almost compresses the maximum data. Transformation mainly involves smoothing, aggregation, generalization and discretization. A. Kusiak et al. introduced new transformation method named feature bundling [7]. When this transformation technique applied to a training data set it embellish classification accuracy of the decision rules generated from this set. Although bundling is destined for integer, categorical and normative features, it can be continued with continues value, for example by using regression function.

Discretization is a process of transforming continuous attribute value into finite set of intervals to generate attributes with a smaller number of distinct values. There are many types of discretization methods such as Direct vs. Incremental, Single vs. Multi attribute, Supervised vs. Unsupervised, Bottom up vs. Top down. For discretization CAIM algorithm is popular [8]. But one of the major drawbacks of CAIM is its stopping criterion, depends on the number of target classes. When the number of target classes is large, its performance drops. C. J. Tsai et al. invented a new discretization algorithm depends on class attribute Contingency Coefficient [9]. They have shown the outperforming results in term of classification accuracy. W. Qu et al. proposed a new Chi2 algorithm named Rectified chi2 algorithm [10], which looks a new merging standard as the basis of interval merging and discretizes the real value attributes absolutely and rationally. J. Ge et al. presented new discretization algorithm for uncertain data which employ both the formula based and sample based probability distribution function [11]. Results showed that algorithm can help the Naive Bayesian Classifier to extent higher classification accuracy. Q. Zhu et al. proposed a novel and effective supervised discretization algorithm based on correlation maximization by using multiple correspondence analysis which is a technique to catch the correlation between feature intervals and classes [12]. The algorithm naturally generates a better set of feature intervals by maximizing their correlations with the classes and gains the classification performance. K. Sriwanna et al. fabricated the Enhanced CAIM (ECAIM) which is the extended version of CAIM [13]. ECAIM is proposed by two modifications. First modification is extended from CAIM to turn in to a real increment discretization method by upgrading the stopping criterion. The second modification is the multi attribute methods by simultaneously considering all attributes instead of a single attribute. Results demonstrate the ECAIM performs better in both synthetic and real world datasets. R. G. Mehta et al. proposed a Modified CAIM (MCAIM) because the results of CAIM are not sufficient in some cases [14]. Intervals generated by MCAIM discretization are more in numbers so some intervals are needed to merge without loss of discretization information. For merging they used CAIR criterion. Experiments showed that MCAIM with merging discretization gives improved results than CAIM and MCAIM without merging discretization methods.

*B. Data Mining Algorithms using Trie Structure*

Trie is an ordered tree data structure that is used to store associative array or dynamic set where the keys are usually strings and the term comes from retrieval [15]. Trie was first proposed by R. Braindais [16]. The most common use of Trie is to represent a set of string [17], [18], [19] which is used in dictionary management. The Trie is also used in different areas such as text compression [20], natural language processing [21], [22], searching for reserved words for a compiler [23], pattern matching [24], [25], IP routing tables [26].

In data mining, visual representation can comfort in strengthening the ability of analysing, understanding the techniques, patterns and their assimilation. But for incremental tree the detail visualization process are rarely introduced. Z. Abdullah et al. have explained the visualization process of constructing the incremental Disorder Support Trie Itemset (DOSTrieIT) data structure from the flat-file dataset. DOSTrieIT is also used as a shrink source of information for building FP Tree [27]. Later year the same author presented a scalable technique to discover items support from Trie data structure [28]. Generally Trie data structure represent frequent patterns via frequent pattern tree (FP Tree). Before the FP Tree can be constructed, there are mainly two scanning process involved in the original database. One of them is to determine the item count (item and their count) that fulfils minimum count threshold by scanning the entire database. However if some changes occurred in the database the process must have find the items and their count again so, Z. Abdullah et al. presented a technique called Fast Determination of Item Support Technique (F-DIST) to capture the items counts from their proposed Disorder Support Trie Itemset (DOSTrieIT) [27] data structure. Experiment said that the computational time to capture the items count using F-DIST from DOSTrieIT is importantly outperformed the FP Tree technique about 3 orders of magnitude for 3 UCI benchmark datasets.

*C. Classification*

Classification maps (or classifies) a data item into one of several predefined categorical classes. Decision tree (DT) can be regarded as a powerful and popular tool for classification and prediction process [29]. DT can be considered more interpretable compared to neural network and support vector machines since they combine more data in an easily understandable format. DT is computationally cheap, easy to use and can deal with uncertainties. It also provides objective analysis to decision making. The drawback of DT is that the whole process requires quantitative data to determine the accuracy of the input.

ID3 (Iterative Dichotomiser 3) is a very simple classification algorithm and based on concept learning ID3, uses information gain to decide which attribute goes into a decision node[30]. The performance of the algorithm degrades as missing values are classified incorrectly. Many researchers proposed modification in the ID3 [31], [32], [33]. C4.5 is the extension of the ID3 but can be used for unavailable values, continuous attribute value range, pruning of the decision tree and rule derivation [34]. The C4.5 generates the decision tree using the information entropy for the classification. Quinlan proposed the C5.0 algorithm which provide improvement over the C4.5. C5.0 is significantly faster, uses memory more efficiently and generates the smaller decision tree than C4.5 [35]. C5.0 supports for boosting which improves the tree and more accuracy. The concept of winnowing is used by C5.0 to automatically winnow the attributes to remove those that may be unhelpful [36].

Schlimmer et al. [37] proposed incremental ID3 algorithm using brute-force method. The drawback of this method is when new data is arrived, an entirely new tree is created. The same author [37] proposed incremental ID4 algorithm. The drawback of this algorithm is that it discards sub trees when new test is chosen for a node. The author Utgoff P.E [38] proposed incremental ID5 which did not discard the sub tree but also not provide guarantee for producing same tree as ID3. The same author [39] Proposed ID5R algorithm. It produces same tree as ID3. It uses recursively updating method for tree's sub nodes. The drawback is that it does not handle numeric values, multicast classification task or missing value. Domingos, Hultern G [40] proposed VFDT (Very Fast Decision Trees) learner reduces the training time for large dataset using subsampling method to incoming data stream. Spencer et al. [41] proposed CVFDT (Concept-adapting Very Fast Decision Trees learner) which can work with concept drift using the sliding window concept on incoming data. The drawback is that it forgot the old data outside the window. G. Fernandes [42] proposed VFDTc which was extension of VFDT for continuous data and concept drift.

III. PROPOSED FRAMEWORK

The stream data is huge in size and with very high dimensionality. As existing data mining algorithms are not suitable for these datasets, the framework for incremental classification is proposed in this paper as shown in the following Fig. 1. The proposed framework contains incremental tree based classification, where the classification model is required to be updated for every unclassified and wrongly classified instances of test and application phase data.

The data containing noise and missing values are to be replaced with suitable data, so the initial stage of the framework deals with preprocessing methods such as missing value replacement, normalization and transformation. The missing value replacement algorithm replaces by mean, median or standard deviation based on data characteristics.

The tree based classification needs categorical data and Trie structured feature tree also deals with categorical data. Discretization is an important stage of the proposed frame work. The preprocessed data is applied to the discretization phase. In this paper MCAIM [14] algorithm is used for discretization which gives better result than previous CAIM algorithm [8]. The prepared data is applied to the feature tree creation phase.

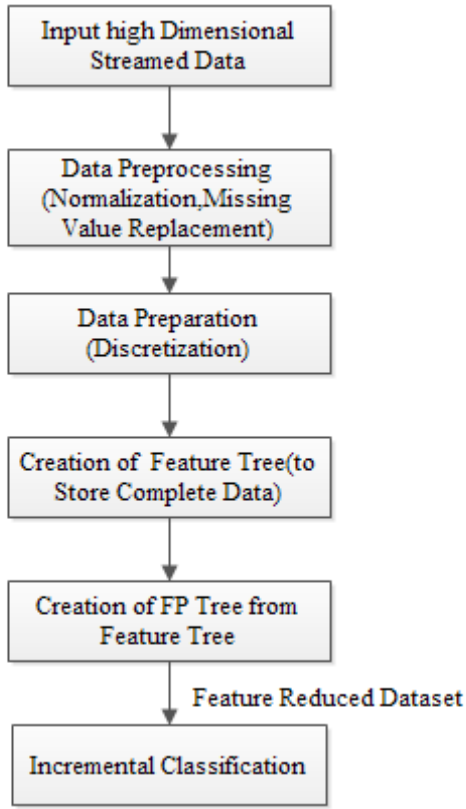

Fig. 1 Block diagram of proposed framework

In classification, as one of the major issue to be dealt with is concept drift. The classification tree can be reconstructed completely in this situation or only affected branch of the tree will be updated. The proposed framework uses updation of the classification tree in this situation, as complete reconstruction will be very time consuming. The storage of the complete data will be a major issue. The Incremental Feature tree (F Tree) is proposed to store the complete historical dataset. Each level of the Trie represents one attribute of the dataset, along with count of every value of the attribute. The complexity of the complete Trie structure will remain same inspite of the varying sized input data.

To improve the efficiency of the classification, the features are required to be reduced as it is a known fact that model tree will not contain every features. Feature reduction phased of the proposed framework uses FP growth algorithm to reduce the features. Trie structured FP tree is used to support incremental FP growth algorithm. The reduced featured dataset is applied to the incremental classification phase. For incremental classification M. Lad et al. proposed ID3 algorithm with combination of CAIM and CAIR for attribute selection criteria [43]. This algorithm is to be updated to update the subtree of the classification tree. The proposed framework allows the classification model to be updated for every unclassified and wrongly classified instances of test and application phase data.

In CVFDT [41] classification algorithm, the window of recent data is managed based on which the classification tree will be updated in case of concept drift. As storage is a major issue for the dataset, it restricts the size of window. The proposed model uses Trie structured feature tree that will reduce this restriction up to remarkable level. The analysis of the classification model, updation in the proposed databases is expected to be performed on the background will not affect the response time of the system. The front end classification tree will be replaced with the background tree in case of updation.

IV. SUMMARY AND FUTURE SCOPE

Stream data has several issues such as high dimensionality, size of the database, and online streaming. The proposed framework makes use of Trie structured feature tree to store the dataset with reduced complexity to maintain the online streamed data. The incremental FP Growth algorithm, in support with Novel incremental classification algorithm will support the online response for the classification process without sacrificing the classification accuracy.

ACKNOWLEDGEMENT

We would like to take this opportunity to convey our sincere gratitude to all those intellectuals concerned for their magnanimous vision by virtue of which we have been guided through to accomplish our mission. We express our heartiest gratitude to Prof. R. G. Mehta and Prof. D. P. Rana for their valuable guidance, moral support and believing in us.